# Non-memoryless Analog Network Coding in Two-Way Relay Channel

Shengli Zhang[1], Soung-Chang Liew[2], Qingfeng Zhou[3], Lu Lu[2], Hui Wang[1]

[1]Department of Communicaton Engineering, Shenzhen University, China
[2]Department of Information Engineering, Chinese University of Hong Kong
[3]Department of Electronic and Computer Engineering, Hong Kong University of Science and Technology
{zsl, wanghsz} @szu.edu.cn, {soung, ll007}@ie.cuhk.edu.hk, enqfzhou@eie.polyu.edu.hk

**Abstract:** Physical-layer Network Coding (PNC) can significantly improve the throughput of two-way relay channels. An interesting variant of PNC is Analog Network Coding (ANC). Almost all ANC schemes proposed to date, however, operate in a symbol by symbol manner (memoryless) and cannot exploit the redundant information in channel-coded packets to enhance performance. This paper proposes a non-memoryless ANC scheme. In particular, we design a soft-input soft-output decoder for the relay node to process the superimposed packets from the two end nodes to yield an estimated MMSE packet for forwarding back to the end nodes. Our decoder takes into account the correlation among different symbols in the packets due to channel coding, and provides significantly improved MSE performance. Our analysis shows that the SNR improvement at the relay node is lower bounded by $1/R$ ($R$ is the code rate) with the simplest LDPC code (repeat code). The SNR improvement is also verified by numerical simulation with LDPC code. Our results indicate that LDPC codes of different degrees are preferred in different SNR regions. Generally speaking, smaller degrees are preferred for lower SNRs.

## I. Introduction

Physical layer network coding (PNC) [1] is a promising technique to improve the throughput of a two-way relay channel (TWRC), in which two end nodes exchange information via a relay node. In PNC, the two end nodes send packets simultaneously to the relay node. The relay node then transforms the superimposed packets to a network-coded packet for broadcast back to the end nodes. Each end node then uses their self information to extract the packet of the other end nodes from the network-coded packet.

Table 1. Classification of PNC schemes according to the processing at the relay node

|  | Memoryless Relay | Non-memoryless Relay |
|---|---|---|
| Finite Field | e.g. PNC [1, 2, 5] | e.g. Coded PNC [7, 8]c |
| Infinite Field | e.g. ANC [5, 6, 9] | ? |

Beyond the above set-up, there are difference variants of PNC schemes. We could classify different PNC schemes into four categories, as in Table. 1. First, the schemes can be classified into memoryless relay and non-memoryless relay, according to whether symbol-by-symbol (memoryless) or packet-by-packet (non-memoryless) processing is performed at the relay. In our paper, non-memoryless relay exploits the channel decoding process to enhance the estimate of the orginal packet; however, it is not necessary to correctly decode the packets as in the traditional decode-and-forward scheme [10]. Generally speaking, memoryless relay schemes are simpler to implement. However, for channel-coded packets, memoryless relay schemes do not make use of the correlations among symbols to remove corruptions due to noise. As a result, noise can accumulate in a multi-hop network with multiple relays. By contrast, the non-memoryless schemes can overcome noise accumulation.

Second, PNC schemes can be classified into finite-field PNC (PNCF) and infinite-field PNC (PNCI) [5] according to the field over which the network coding at the relay operates. Generally speaking, PNCF generates less extraneous information at the relay and is more efficient for downlink transmission; while PNCI can match the two uplink channels to reduce estimation errors.

Since different schemes are preferred in different scenarios, most of them are of interest and have been studied to a certain extent. However, to the best of our knowledge, there have been no proposals or investigations on non-memoryless PNCI schemes, such as Analog Network Coding (ANC). The reasons could be that: i) it is not straightforward for the relay to decode the received packet $h_{1,3}X_1+h_{2,3}X_2$ ($X_1$ and $X_2$ are the packets from the two end nodes) since it is not a valid codeword; ii) for most channels, the achievable end-to-end code rates are beyond the capacity of the relay node, which does not have self-information of the end nodes. The situation faced by the relay in a PNC system, therefore, is different from that in traditional channel coding, which focuses on code rates within the capacity region[1].

In this paper, we propose a novel channel decoding (non-memoryless) scheme at the relay node to enhance the performance of ANC. Note that with ANC, we do not aim to successfully decode $X_1$, $X_2$, or $X_1 \oplus X_2$, but to find a good estimate for $h_{1,3}X_1+h_{2,3}X_2$. We assume the same LDPC code is used at both end nodes. First, the joint probability of the two symbols from the end nodes is estimated from the received superimposed symbol. Exploiting the correlations among the coded symbols, this joint probability is then refined by a novel belief propagation algorithm [12]. Based on the refined joint probability, an MMSE estimate of $h_{1,3}X_1+h_{2,3}X_2$ is obtained as the network-coded signals for broadcast back to the two end nodes. As shown in [5, 11], MMSE estimate at the relay achieves better power allocation to symbols with varying reliabilities; and achieves smaller mean square uncorrelated error (MSUE) at the end node than any other forms of estimate for $h_{1,3}X_1+h_{2,3}X_2$. We analyze the performance of our scheme in terms of the SNR improvement for a given channel code rate. An achievable lower bound is given.

---

[1] Recently, there have been several works on decoding channel code with beyond-capacity rate for point-to-point channel [10, 11, 17].

## II. System Model and Notations

*System:*

We consider the two-way relay channel as shown in Fig.1, in which nodes $N_1$ and $N_2$ exchange information with the help of relay node $N_3$. We assume that all nodes are half-duplex, i.e., a node cannot receive and transmit simultaneously. This is an assumption arising from practical considerations because it is difficult for the wireless nodes to remove the strong interference of its own transmitting signal from the received signal. We also assume that there is no direct link between nodes $N_1$ and $N_2$. An example in practice is a satellite communication system in which the two end nodes on the earth can only communicate with each other via the relay satellite.

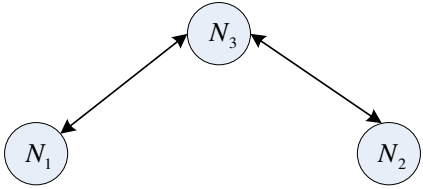

Fig 1: Two-way relay channel.

In this paper, $S_i$ is a vector denoting the uncoded source packet of node $N_i$; $X_i$ denotes the packet after channel coding; $A_i$ denotes the transmitted packets after BPSK modulation; and $Y_i$ denotes the received base-band packet at node $N_i$. Lowercase letters, $s_i \in \{0,1\}$, $a_i \in \{-1,1\}$, $x_i \in \{0,1\}$, or $y_i \in \mathbb{R}$, denote a symbol in the corresponding packet. The complex channel coefficient from node $N_i$ to node $N_j$, $h_{i,j}$, is assumed to be invariant during one packet transmission, and varies independently between different packets.

The two-phase transmission scheme in Physical layer Network Coding (PNC) consists of an uplink phase and a downlink phase. In the uplink phase, $N_1$ and $N_2$ transmit to $N_3$ simultaneously. Therefore, $N_3$ receives

$$y'_3 = h_{1,3}a_1 + h_{2,3}a_2 + w'_3 \\ = h_{1,3}(1-2x_1) + h_{2,3}(1-2x_2) + w'_3 \quad (1)$$

where $w'_3$ is the noise at $N_3$, which is complex Gaussian with variance $\sigma^2$ (identical for all the three nodes), We assume that the transmit power, the phase difference between the transmitted signal and the local signal at the receive node, and the channel fading effect at the received node $N_j$ are taken into account by $h_{i,j}$, which is assumed to be complex Gaussian variable with a given variance.

With soft decision demodulation, the received signal at $N_3$ can be expressed as

$$y_3 = -(y'_3 - h_{1,3} - h_{2,3}) = 2h_{1,3}x_1 + 2h_{2,3}x_2 + w_3 \quad (2)$$

where the Gaussian noise $w_3 = -w'_3 \in CN(0,\sigma^2)$ and its vector version for the overall packet is $W_3$. Hereafter, we write the received packet $Y_3$ as a function of the transmitted packet $h_{1,3}X_1 + h_{2,3}X_2$.

In the downlink phase, a memoryless system would just receive a superimposed symbol $y_3$, process it and broadcast it to both end nodes. For example, the ANC scheme [9] simply amplifies $y_3$ by a fixed factor and broadcasts $x_3 = \alpha y_3$, where $\alpha$ is a scaling factor to satisfy the power constraint. Indeed, $x_3$ is a linear MMSE estimation of $h_{1,3}x_1 + h_{2,3}x_2$. We can write the signals received by $N_1$ and $N_2$ as

$$y_1 = h_{3,1}x_3 + w_1 \qquad y_2 = h_{3,2}x_3 + w_2 \quad (3)$$

Consider $N_1$. It obtains its target information by subtracting the self-information as

$$y_1' = y_1 - h_{3,1}\alpha h_{1,3}x_1 \\ = h_{3,1}\alpha h_{2,3}x_2 + h_{3,1}\alpha w + w_1 \quad (4)$$

The above equation is of the same form as that in a point-to-point transmission system: it consists of the target signal plus noise. Thus, $N_1$ could decode $S_2$ as in the point-to-point system.

Examples of other memoryless schemes are [5, 6]. This paper is different from these previous works in that it proposes to use non-memoryless estimation to exploit the correlations among the symbols in a channel-coded packet. In this case, the estimate of each symbol $h_{1,3}x_1 + h_{2,3}x_2$ depends on the whole received packet, and it can be expressed as

$$x_3 = func(Y_3) . \quad (5)$$

*LDPC codes:*

An integral part of our new signal processing scheme at the relay is in fact a channel decoding scheme, except that we do not aim to always successfully decode the individual packets at the relay. The channel decoding scheme depends on the channel codes adopted at the end nodes. For simplicity, we assume the same LDPC code [14] is used at the end nodes. LDPC code is attractive in that it is capacity approaching [15]. An LDPC code can be characterized by a sparse parity check matrix H. Suppose the uncoded packet length is $N-K$ and the coded packet length is $N$. Denote the $(N-K) \times N$ parity check matrix by $H$ and the corresponding $N \times (N-K)$ generator matrix by $G$. Then we have

$$X_i = \Gamma(S_i) = GS_i \quad HX_i = 0 . \quad (6)$$

and the code rate is $1-K/N$.

## III. Channel Decoding Algorithm

This section elaborates the proposed channel decoding scheme at the relay. Although we focus on regular LDPC code in this paper, extensions to other channel codes, such as RA code and Turbo code, are straightforward.

At the relay, the target is to obtain a refined estimate of

$h_{1,3}X_1 + h_{2,3}X_2$ based on the received packet $Y_3$ and the redundancy contained in the channel-coded packets, $X_1$ and $X_2$. To do so, we use belief propagation to decode the joint probability density functions of the 2-tuple $x=(x_1, x_2)$, denoted by $P(x_1, x_2)$, from $Y_3$. In order to perform channel decoding, we regard the 2N-tuple $(X_1, X_2)$ as one virtual code and the corresponding vector version constraint is

$$HX = H(X_1, X_2) = (HX_1, HX_2) = (0,0). \qquad (7)$$

Consider one check node which is connected by $k$ edges in Fig. 2. The above constraint can be expressed in scalar form as

$$g(x[1], x[2], \cdots x[k]) = (\sum_{i=1}^{k} x_1[i], \sum_{i=1}^{k} x_2[i]) = (0,0) \qquad (8)$$

From the above equation, the virtual code is equivalent to a 4-ary LDPC code, and its corresponding Tanner Graph is shown in Fig. 2. The belief propagation decoding algorithm can then be designed accordingly.

Let $P_{h,t}$ denote the message passed between a check node and a variable node (code node). The message is associated with the edge from node $h$ to node $t$, where one of $h$ or $t$ is a variable node, and the other is a check node. Let $P_k$, $k \in [1,N]$, be the message from the $k$-th (ordered from top to bottom as in Fig. 2) evidence node to the $k$-th code node, where $N$ is the length of the coded packet.

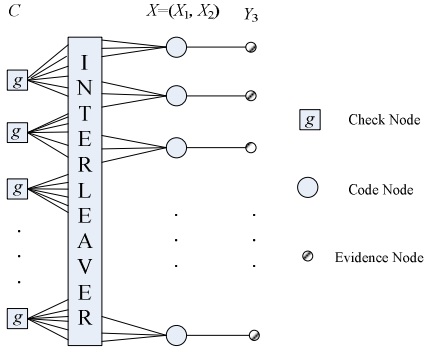

Fig 2: Tanner Graph of the virtual (3, 6) LDPC code

*Message form:*

$P_{h,t} = (p_{0,0}, p_{0,1}, p_{1,0}, p_{1,1})$ is a vector, in which $p_{i,j}$ is the probability that the corresponding variable node ($h$ or $t$) takes on the pair of values of $(i, j)$ $i, j \in \{0,1\}$.

$P_k = (p_{0,0}, p_{0,1}, p_{1,0}, p_{1,1})$ is a vector, in which $p_{i,j}$ is the probability that the $k$th coded symbol is $(i, j)$ given the $k$-th received symbol.

*Message Initial Values*:

All the messages associated with the edges in Fig. 2 are initialized to (1/4, 1/4, 1/4, 1/4) except for the messages on the edges incident to the evidence nodes, which contain information on the received signal. The message from an evidence node is computed from the corresponding received symbol $y_3$ as follows:

$$\begin{aligned}P_k &= (p_{0,0}, p_{0,1}, p_{1,0}, p_{1,1}) \\ &= \big(\Pr(x_1=0, x_2=0 \mid y_3), \Pr(x_1=0, x_2=1 \mid y_3), \\ &\quad \Pr(x_1=1, x_2=0 \mid y_3), \Pr(x_1=1, x_2=1 \mid y_3)\big) \\ &= \frac{1}{\beta}\bigg(\exp(\frac{-\mid y_3 - h_{1,3} - h_{2,3}\mid^2}{2\sigma^2}), \exp(\frac{-\mid y_3 - h_{1,3} + h_{2,3}\mid^2}{2\sigma^2}), \\ &\quad \exp(\frac{-\mid y_3 + h_{1,3} - h_{2,3}\mid^2}{2\sigma^2}), \exp(\frac{-\mid y_3 + h_{1,3} + h_{2,3}\mid^2}{2\sigma^2})\bigg)\end{aligned} \qquad (9)$$

where $\beta$ is a normalizing factor to make sure that the four probabilities sum to one.

*Message Update Rules*:

Parallel to the generic update rules in [12], we also have the same message update rules at our check nodes and variable nodes. Note that the messages from the evidence nodes to the code nodes remain the same without being changed during the iterations of the decoding process.

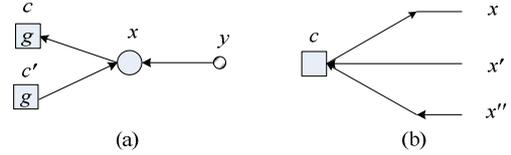

Fig 3: message updates for decoding the virtual code in Fig. 2

*Update Equations for Output Messages Going Out of a Variable Node*

When the node degree is 2, each output message is the same as the other input message.

Fig. 3(a) illustrates the case in which the node degree is 3. When the probability vectors of the two input messages, $P = (p_{0,0}, p_{0,1}, p_{1,0}, p_{1,1})$ and $Q = (q_{0,0}, q_{0,1}, q_{1,0}, q_{1,1})$ (associated with the edge from $y$ to $x$ and the edge from $s'$ to $x$, respectively), arrive at a code node of degree three (except the lowest code node), the probability that the code symbol is (0,0) is obtained as follows:

$$\Pr(x=(0,0) \mid P, Q) = 4\lambda p_{0,0} q_{0,0} \qquad (10)$$

where $\lambda = \Pr(P)\Pr(Q)/\Pr(P,Q)$ and the two input messages are assumed to be independent given the value of the variable node, i.e., $\Pr(P \mid Q, x) = \Pr(P \mid x)$. Given the $l$-depth neighborhood of the edge is cycle free (cycle free condition), this assumption is true for iterations up to $l$ in the decoding algorithm. As in the proof for the LDPC codes in [14], the probability that the cycle free condition is true for our coder in Fig. 2 should also go to 1 as the length of the code goes to infinity. That is, $l$ becomes larger and larger.

In a similar way, we can obtain that $\Pr(x=(0,1) \mid P, Q) = 4\lambda p_{0,1} q_{0,1}$, $\Pr(x=(1,0) \mid P, Q) = 4\lambda p_{1,0} q_{1,0}$,

and $\Pr(x=(1,1)|P,Q) = 4\lambda p_{1,1}q_{1,1}$. Thus, the output message at the variable node is

$$VAR(P,Q) = 4\lambda(p_{0,0}q_{0,0}, p_{0,1}q_{0,1}, p_{1,0}q_{1,0}, p_{1,1}q_{1,1}) \quad (11)$$

Since the summation of the three probabilities should be 1, we require $\lambda = 1/(p_{0,0}q_{0,0} + p_{0,1}q_{0,1} + p_{1,0}q_{1,0} + p_{1,1}q_{1,1})/4$ for normalization.

When the node degree is $k$, and the $k$-1 input messages are $P^1, \cdots P^{k-1}$. Then, we can obtain the output message in an induction way as

$$VAR(P^1, \cdots P^{k-1}) = \lambda(\prod_{i=1}^{k-1} p_{0,0}^i, \prod_{i=0}^{k-1} p_{0,1}^i, \prod_{i=0}^{k-1} p_{1,0}^i, \prod_{i=0}^{k-1} p_{1,1}^i) \quad (12)$$

where $\lambda$ is the normalization factor.

*Update Equations for Output Messages Going Out of Check Nodes:*

When the node degree is 2, each output message is the same as the other input message.

Fig. 3(b) illustrates the case in which the node degree is 3. Based on the $f$ defined in (8), and using similar computation as in (8), the probability that the variable node symbol $x$ is (0,0) given the two input messages $P = (p_{0,0}, p_{0,1}, p_{1,0}, p_{1,1})$ and $Q = (q_{0,0}, q_{0,1}, q_{1,0}, q_{1,1})$ (associated with the edge from $x'$ to $s$ and the edge from $x'$ to $c$, respectively) is

$$\Pr(x=(0,0)|P,Q) = p_{0,0}q_{0,0} + p_{0,1}q_{0,1} + p_{1,0}q_{1,0} + p_{1,1}q_{1,1} = g_{0,0}(P,Q) \quad (13)$$

In a similar way, we can obtain that $\Pr(x=(0,1)|P,Q)$, $\Pr(x=(1,0)|P,Q)$ and $\Pr(x=(1,1)|P,Q)$, which are denoted by $g_{0,1}(P,Q)$, $g_{1,0}(P,Q)$, $g_{1,1}(P,Q)$ respectively. As a result, the output message at the check node is

$$\begin{aligned}&CHK(P,Q)\\&= (g_{0,0}(P,Q), g_{0,1}(P,Q), g_{1,0}(P,Q), g_{1,1}(P,Q)) = g(P,Q)\end{aligned} \quad (14)$$

When the node degree is $k$ and the $k$-1 input messages to the variable node are $P^1, \cdots P^{k-1}$. Then, we can obtain the output message in a recursive manner as

$$CHK(P^1, \cdots, P^{k-1}) = g(\cdots g(g(P^1, P^2), P^3) \cdots P^{k-1}) \quad (15)$$

*Stop Rules*:

The messages are updated in an iterative way and the iteration stops when the following rules are given. We first check whether the decoding of $X_1$ is successful. We make a hard decision on $x_1$ by taking the marginal probability:

$$\hat{x}_1 = \begin{cases} 0 & \text{if } p_{0,0} + p_{0,1} \geq p_{1,0} + p_{1,1} \\ 1 & \text{otherwise} \end{cases} \quad (16)$$

If $H\hat{X}_1 = 0$, then the decode of $X_1$ is successful and we stop its iteration by setting the messages associated with each edge as

$$P = \begin{cases} (p_{0,0}+p_{1,0}, p_{0,1}+p_{1,1}, 0, 0) & \text{if } \hat{x}_1 = 0 \\ (0, 0, p_{0,0}+p_{1,0}, p_{0,1}+p_{1,1}) & \text{if } \hat{x}_1 = 1 \end{cases}. \quad (17)$$

It can be verified that the value of $\hat{X}_1$ will not change any more with the message update rules under the setting in (17). This is equivalent to subtracting the interference of $X_1$ from $Y$ and decoding $X_2$ alone.

Similarly, we can make a hard decision on $X_2$ and check whether it has been correctly decoded.

If both $X_1$ and $X_2$ have been correctly decoded, the iteration stops. However, in general, it may not always be possible to decode the packets successfully. So we need to set a maximum number of iterations.

*The Relay Output*:

When the decoding process stops, the relay will generate the output symbols based on the output of the decoder. If the number of iterations is less than the maximum value, then both packets have been correctly decoded, and we make a hard decision on the probability tuple. Finally, the $k$-th symbol to be broadcast is generated by

$$x_3[k] = \sum_{i,j \in \{0,1\}} p_{i,j}[k] \times [(1-2i)a + (1-2j)b]. \quad (18)$$

In the above equation, we simply set $(a, b)$ to $(h_{1,3}, h_{2,3})$ to match the uplink channel[2]. Then, we can regard the broadcast symbol as

$$x_3 = E\{h_{1,3}x_1 + h_{2,3}x_2 | Y_3\}. \quad (19)$$

In other words, $x_3$ is an MMSE estimate based on the observation of the whole packet. This contrasts with the memoryless PNCI MMSE estimate given by $x_3 = E\{h_{1,3}x_1 + h_{2,3}x_2 | y_3\}$.

## IV.     Performance Analysis

MMSE relay is a good relay scheme [5, 11] which can minimize the MSUE and achieve smaller BER at the end nodes. MMSE is closely related to uncoded BER and mutual information [16]. In this section, we analyze the performance of the proposed non-memoryless ANC scheme in terms of minimum mean square error (MMSE) at the relay node. The MSE of the proposed scheme is defined as

$$mse = E\{|x_3 - h_{1,3}x_1 - h_{2,3}x_2|^2\}. \quad (20)$$

---

[2] Besides pure real-field network coding described above, a hybrid PNC is also possible: if the two end packets can be decoded, the broadcast packet could be $X_3 = X_1 \oplus X_2$; otherwise, it is $X_3$ as given by (18) (an indicator in the packet header can be used to indicate which has been sent).   Even for our real-field network coding form as in (18), there are more possible values for $(a, b)$. This consideration will be addressed in future work.

For comparison, we also present the MMSE of the "conventional" memoryless (un-decoded) ANC given in [5]. The MMSE estimate of the received signal is [5]

$$x_3' = \sum_{i,j} p_{i,j}(h_{1,3}(1-2i) + h_{2,3}(1-2j)) \quad (21)$$

where $p_{i,j}$ takes on the values in (9). The corresponding MSE is

$$mse\_con = E\{|x_3' - h_{1,3}x_1 - h_{2,3}x_2|^2\} \quad (22)$$

When the channel coefficients are fixed, $mse\_con$ is a function of the Gaussian noise variance $\sigma^2$, and we denote it by

$$mse\_con = f_1(\sigma^2) \quad (23)$$

We now investigate the SNR improvement, which is defined as the extra SNR needed by the memoryless scheme to achieve the same MSE as our non-memoryless scheme. Specifically, the SNR improvement is

$$\Delta SNR = snr\_con - snr = 10\log \frac{f_1^{-1}(mse)}{f_1^{-1}(mse\_con)}. \quad (24)$$

*Lower bound:*

We first present a lower bound of the SNR improvement. Consider the naïve repeat channel code in which each symbol is repeated $q$ times, i.e., the $(1, q)$ LDPC code. The relay node may combine the same symbols with maximum ratio combination (MRC)[3] and the resulting SNR is raised by a factor of $q$. Therefore, the SNR improvement in (24) is also $10\log(q)$, with respect to the memoryless ANC scheme. This result can be easily extended to the case of non-integer repeat factor $q$ by repeating some bits $\lfloor q \rfloor$ times and other bits $\lfloor q \rfloor +1$ times. Since repeat code is one specific LDPC code, we have Proposition 1 as follows:

*Proposition 1:* For a channel code with rate $R$, the SNR improvement is lower bounded by

$$\Delta SNR \geq -10\log(R). \quad (25)$$

The simple repeat code works well in low SNR. When the channel is good, the more sophisticated LDPC codes perform better and the lower bound $-10\log(R)$ is not tight in general.

## V. Numerical Simulation

This section gives numerical simulation results to show the performance improvement of the proposed non-memoryless ANC scheme.

In our simulation, the parity check matrix $H$ of the regular LDPC code is randomly generated according to the Gallager's method. The length of the coded packet $N$=1800, while three different column degree and row degree pairs, (3, 6), (2, 4) and (1, 2), are explored to investigate the performance of non-memoryless ANC under different channel codes (fixed rate 0.5). The $(1, 2)$ case in fact corresponds to a naïve repeat code. The maximum number of decoding iterations is set to 20. As in (9), the channel decoding only depends on the distance between different constellation points of $h_{1,3}x_1 + h_{2,3}x_2$. Here we simulate the performance improvement of non-memoryless ANC under different distances among the constellation points and simply set $h_{1,3}$ and $h_{2,3}$ to real values.. The SNR in our simulation is defined as $1/\sigma^2$, where $\sigma$ is the noise variance at the relay node.

---
[3] The BP algorithm is equivalent to MRC for the $(1, 2)$ LDPC code.

In Fig. 4, the MSE at the relay node is an average over 100 packets and the parity check matrix $H$ is regenerated for each packet in a random way for non-memoryless and memoryless ANC with the channel coefficients of $h_{1,3} = h_{2,3} = 1$. As shown in the figure, the $(1, 2)$ LDPC code performs best and the $(3, 6)$ LDPC code performs worst when the SNR is less then 2.5 dB. When the SNR is more than 2.5dB, the $(3, 6)$ LDPC code becomes the best while the $(1, 2)$ LDPC code is the worst. For the SNR region being simulated, the BER at the relay node is always about 0.2, since the decoder can not differentiate the two tuples $(1, -1)$ and $(-1, 1)$ when 0 is received. However, this is not important because we need at the relay is the estimate for $h_{1,3}x_1 + h_{2,3}x_2 = x_1 + x_2$, not $(x_1, x_2)$.

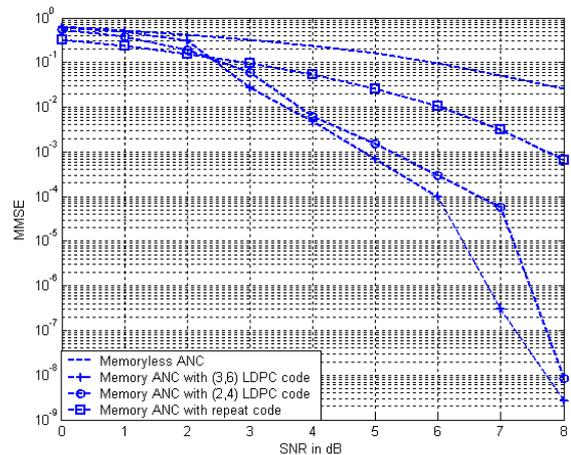

Fig 4: MMSE performance when $h_{1,3}$=1 and $h_{2,3}$=1

In Fig. 5, we simulate the MMSE performance for random channel coefficients. We simulate 1000 packets to obtain the average performance and the channel coefficients $h_{1,3}$ and $h_{2,3}$ are randomly generated for each packet with Rayleigh distribution. We can see that the performance of all the three coded schemes degrade with Rayleigh distributed channel coefficients. However, the relative performance is the same as in the previous simulation.

All the simulation results show that the complex LDPC codes are good at distinguishing the constellation points spread far apart, while they are bad at distinguishing compact constellation points. The simple repeat code works in the opposite way. The intuition is as follows. In low SNR region, there is a high probability that in a complex code, a check node is connected to two or more very poorly received symbols, and that there is one or more bad symbols participating in the computation of (13) for a variable node. The uncertainty in variable nodes propagates to other variable nodes in the BP inferencing process under a complex LDPC code. In contrast, the simpler LDPC codes perform well because the variable nodes are not as intertwined together. In high SNR region, there is a small probability that the check node connects to two very bad symbols. In this case, the certainty in variable nodes with good symbols propagates to other variables nodes.

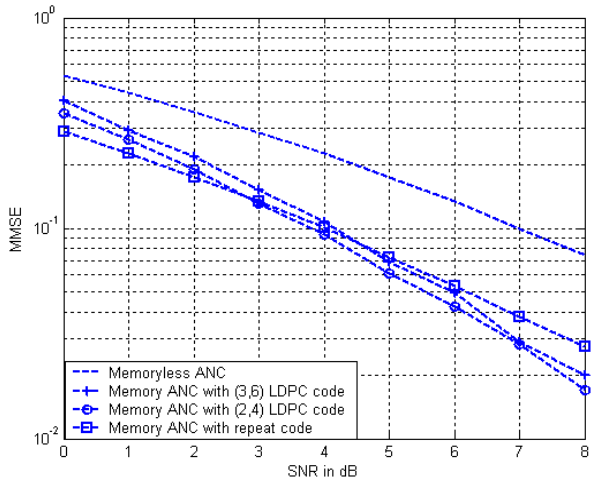

Fig 5: MMSE performance when $h_{1,3}$, $h_{2,3}$ are Rayleigh distributed random variables with unit variance

## VI. Conclusion

In this paper, we proposed a memory amplify-and-forward network coding relay scheme for two way relay channels. Specifically, we propose a new soft input soft output decoding algorithm to refine the estimate of superimposed signal at the relay node. Our analysis showed that the new scheme can improve the SNR at the relay by 1/R at least with repeat code. Our numerical simulation shows that the new scheme employing LDPC codes can improve the MMSE at the relay node by at least 3dB and at most 6dB. The SNR improvement at the relay node translates to better uplink channels, which improves the performance at the end node ultimately. More interestingly, we find that LDPC codes with different complexities (the number of 1's in the parity check matrix *H*) perform best in different SNR regions. In low SNR region, the simplest LDPC code (repeat code) performs best; in high SNR region, LDPC codes with large degrees perform best; in middle SNR region, the LDPC code with moderate degrees performs best.

Reference:

[1]. S. Zhang, S. Liew, and P. Lam, "Physical layer network coding," Mobicom2006, LA, 2006.
[2]. W. Nam, S. Y. Chung, and Y. H. Lee, "Capacity Bounds for Two-Way Relay Channels", IEEE International Zurich Seminar on Communications, 2008.
[3]. C. Fragouli, and E. Soljanin, "Network coding applications", Foundations and Trends in Networking, 2007
[4]. B. Nazer, and M. Gastpar, "Lattice coding increases multicast rates for Gaussian multiple-access networks" 45th Annual Allerton Conference, 2007.
[5]. S. Zhang, S. C. Liew, and L. Lu, "Physical layer network coding schemes over finite and infinite fields", *Proc. IEEE Globecom*, 2008.
[6]. T. Cui, T. Ho, and J. Kliewer, "Memoryless Relay Strategies for Two-Way Relay Channels: Performance Analysis and Optimization" *Proc. IEEE ICC* 2008.
[7]. S. Zhang and S.-C. Liew, "Channel coding and decoding in a relay system operated with physical-layer network coding," *IEEE Journal on Selected Areas in Communications*, vol. 27, no. 5, pp. 788–796, 2009.
[8]. T. Koike-Akino, P. Popovski, and V. Tarokh, "Denoising strategy for convolutionally–coded bidirectional relaying," *Proc*. of *IEEE ICC* 2009, Dresden, Germany, June 2009.
[9]. S. Katti, S. Gollakota, and D. Katabi, "Embracing wireless interference: Analog network coding", Proceedings of SigComm 2007.
[10]. X. Bao and J. Li, "Efficient Message Relaying for Wireless User Cooperation: Decode-Amplify- Forward (DAF) and Hybrid DAF and Coded-Cooperation," IEEE Transaction on Wireless Communications, vol. 6, no. 11, pp. 3975–3984, Nov. 2007.
[11]. P. Weitkemper, D. Wubben, and K.-D. Kammeyer, "Minimum MSE Relaying in Coded Networks," in International ITG Workshop on Smart Antennas (WSA' 08), Darmstadt, Germany, Feb. 2008.
[12]. F. R. Kschischang, B. J. Frey, H.-A. Loeliger, "Factor graphs and the sum-product algorithm," *IEEE Trans. on Information Theory*, vol. 47, no. 2, Feb. 2001.
[13]. MIT sigcom
[14]. R. G. Gallager, "Low Density Parity-Check Codes." MIT Press, Cambridge, MA, 1963.
[15]. D. J. C. MacKay and R. M. Neal, "Near Shannon Limit Performance of Low Density Parity Check Codes," Electronics Letters, July 1996.
[16]. D. Guo, S. Shamai, and S. Verdú, "Mutual information and minimum mean-square error in Gaussian channels," IEEE Trans. Inf. Theory, vol. 51, no. 4, pp. 1261–1282, Apr. 2005.
[17]. A. Bennatan, A. B. Calderbank, and S. Shamai, "Bounds on the MMSE of bad LDPC codes at rates above capacity", in Proc. 46-Allerton, Sep. 2008.